\documentclass[prl,aps,twocolumn,showpacs,amsmath,amssymb,floatfix]{revtex4}

\usepackage[dvips]{graphicx}
\usepackage{bm}

\newcommand{\be}{\begin{equation}}
\newcommand{\ee}{\end{equation}}
\newcommand{\grad}{\bm\nabla}
\newcommand{\curl}{\bm\nabla\wedge}
\newcommand{\divr}{\bm\nabla\cdot}

\begin{document}

\title{Vortex lattices in Bose-Einstein condensates:\\
from the Thomas-Fermi to the lowest Landau level regime}
\author{M.~Cozzini}
\author{S.~Stringari}
\author{C.~Tozzo}
\affiliation{Dipartimento di Fisica, Universit\`a di Trento and BEC-INFM,
I-38050 Povo, Italy}

\date{\today}

\begin{abstract}

We consider a periodic vortex lattice in a rotating Bose-Einstein condensed gas,
where the centrifugal potential is exactly compensated by the external harmonic
trap.
By introducing a gauge transformation which makes the Hamiltonian periodic, we
solve numerically the 2D Gross-Pitaevskii equation finding the exact mean field
ground state.
In particular, we explore the crossover between the Thomas-Fermi regime,
holding for large values of the coupling constant, and the lowest Landau level
limit, corresponding to the weakly interacting case.
Explicit results are given for the equation of state, the vortex core size, as
well as the elastic shear modulus, which is crucial for the calculation of the
Tkachenko frequencies.

\end{abstract}

\pacs{03.75.Kk, 03.75.Lm, 67.40.Vs}

\maketitle

In the last years, a significant effort has been devoted to the study of vortex
lattices in harmonically trapped rotating Bose-Einstein condensates.
Striking results have been obtained experimentally, leading to the observation
of large vortex lattices in fast rotating condensates \cite{exp vort} and to
the measurement of their dynamical properties \cite{JILA stripes,JILA tka}.
Theoretically, several predictions have been provided \cite{TN stripes,B-P,M-H
spin} in agreement with experiments and more challenging regimes, related to
the quantum Hall effect, have also been proposed \cite{QHE}. 

\begin{figure}
\includegraphics[width=8.5cm]{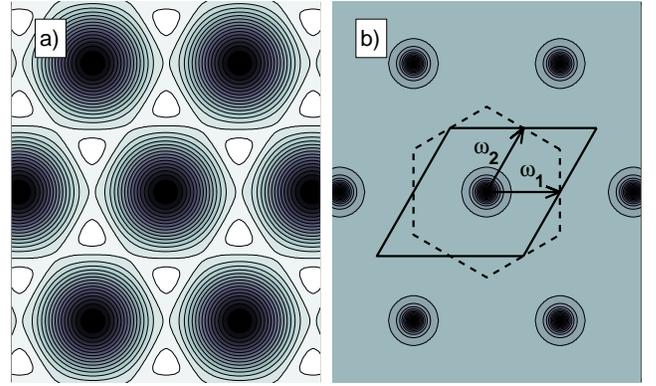}
\caption{\label{fig:1}%
Density distribution of the condensate in the $x$-$y$ plane  computed by
solving Eq.~({\ref{eq:periodicGP}}) for
(a) $g=0$, (b) $g=300$.
In (b) the dashed hexagon is the Wigner-Seitz cell, while the solid
parallelogram is the computation box (see text).
Darker regions correspond to lower density.}
\end{figure}

In this paper we study the rotating analogue of a uniform Bose-Einstein
condensate, where a periodic vortex lattice is present. For dilute atomic
gases, which are highly compressible, this can only be obtained when the
centrifugal potential is exactly cancelled by the harmonic confinement, i.e.,
when the harmonic trapping frequency is precisely equal to the angular velocity
$\Omega$.
Under this condition, in spite of the non-periodicity of the Gross-Pitaevskii
(GP) equation governing the system, the density $n$ (see Fig.~\ref{fig:1}) and
the velocity field $\bm{v}'$ in the rotating frame are periodic in the plane of
rotation \cite{F-S-S,tka}, as can be proved by a proper gauge transformation.
Analytical predictions are available in the literature for both the weakly
interacting regime \cite{Ho LLL,M-H spin,C2Sonin,aftalion,watanabe},
where the system is in the
lowest Landau level (LLL), and the strongly interacting one \cite{tka}, where
the density is basically constant and the Thomas-Fermi (TF) approximation
applies.
Solving numerically the GP equation, we recover these limiting cases and
compute the ground state for any value of the interaction strength.
The validity of the mean-field approach requires that the number of atoms per
vortex be very large. If this condition is not satisfied the system enters
a strongly correlated regime related to the quantum Hall effect \cite{QHE}.

In the rotating frame, the external confinement $V_{\text{ext}}$ combines with
the centrifugal potential, giving rise to the effective trapping
$V_{\text{eff}}=V_{\text{ext}}-m\Omega^2(x^2+y^2)/2$, where $m$ is the atomic
mass. As anticipated above, we impose the compensation $V_{\text{eff}}=0$ and
decouple the motion in the axial direction from the
radial one, therefore reducing to a 2-dimensional problem with an effective
coupling constant $g_{2D}$ \cite{2D}.
In addition, our discussion can be conveniently reformulated in terms of
dimensionless quantities by using harmonic oscillator units, i.e.,
$\hbar\Omega$, $\Omega$, and $l_\Omega=\sqrt{\hbar/m\Omega}$ for energy,
frequency, and length respectively. The length $l_\Omega$
also fixes the equilibrium inter-vortex distance, according to the Feynman
expression for the vortex density $n_v=m\Omega/\pi\hbar=1/\pi{l_\Omega^2}$
\cite{Feynman}. Having to deal with an infinite periodic vortex lattice, we fix
the average particle density $\langle{n}\rangle$ in terms of the number
$N_{\text{cell}}$ of atoms per lattice cell,
$\langle{n}\rangle=N_{\text{cell}}/A_{\text{cell}}$, where
$A_{\text{cell}}=1/n_v=\pi$ is the cell area.
The corresponding dimensionless stationary Gross-Pitaevskii equation in the
rotating frame is hence
\be \label{eq:GP}
\mu\psi = \left[
\frac{(-i\grad-\bm{e}_z\wedge\bm{r})^2}{2}+g|\psi|^2\right]\psi \ ,
\ee
where $\mu$ is the chemical potential, $\bm{e}_z$ is the unit vector along $z$,
and the dimensionless coupling constant 
$g={\pi}g_{2D}\langle{n}\rangle/\hbar\Omega=g_{2D}N_{\text{cell}}m/\hbar^2$
is the only parameter governing the equation. The normalization
is $\int|\psi|^2=1$, where $\int\equiv\int_{A_{\text{cell}}}\text{d}\bm{r}$
means integration over a single cell.

Due to the presence of the rotational term $\bm{e}_z\wedge\bm{r}$,
Eq.~(\ref{eq:GP}) is not spatially periodic.
However, in the spirit of the magnetic field analogy discussed in
Refs.~\cite{QHE,Ho LLL}, one can perform a gauge transformation
which turns the symmetric effective vector
potential $\bm{A}=\bm{e}_z\wedge\bm{r}$ into a periodic function
$\bm{A}'=\bm{A}-\grad\Lambda$.
We choose $\Lambda=S_T$, where $S_T$ is the (non-periodic) phase associated
with the Tkachenko (non-periodic) velocity field $\bm{v}_T=\grad{S}_T$ used by
Tkachenko to describe the vortex lattice of an incompressible fluid \cite{tka}.
With such a choice, indeed, one finds
%\be \label{eq:gauge}
$
\bm{A}'=\bm{e}_z\wedge\bm{r}-\bm{v}_T=-\bm{v}_T'\;,
$
%\ee
where the Tkachenko
velocity $\bm{v}_T'$ in the \textit{rotating frame} is now periodic \cite{tka}.
We therefore write $\psi=\tilde\psi e^{iS_T}$ and
substitute this expression into Eq.~(\ref{eq:GP}), finally obtaining
\be \label{eq:periodicGP}
\mu\tilde\psi =
\left[\frac{(-i\grad+\bm{v}'_T)^2}{2}+g|\tilde\psi|^2\right]\tilde\psi \ ,
\ee
which admits periodic solutions.
The explicit expression of $\bm{v}_T$ is conveniently written using the
complex variable representation $z=x+iy$, $v=v_x+iv_y$ \cite{complex}.
The lattice geometry is determined by the half-periods $\omega_1$ and
$\omega_2$ (see Fig.~\ref{fig:1}(b))
which, due to the constraint $A_{\text{cell}}=\pi$, are related by
$\text{Im}(\omega_1^*\omega_2)=\pi/4$.
One then has $v_T(z^*)=i[\zeta(z)+\alpha{z}]^*$,
where $\zeta(z)\equiv\zeta(z;\omega_1,\omega_2)=
1/z+\sum_{jk}'[1/(z-z_{jk})+1/z_{jk}+z/z_{jk}^2]$ is the
\textit{quasi-periodic} Weierstrass $\zeta$-function \cite{chandra}. Here the
primed summation excludes the term $j\!=\!k\!=\!0$, $z_{jk}=2j\omega_1+2k\omega_2$
are the vortex positions, and
$\alpha=[\omega_1^*-\zeta(\omega_1)]/\omega_1$ is chosen to
guarantee the periodicity of $v_T'(z,z^*)=v_T(z^*)-iz$.
Correspondingly, the Tkachenko phase is given by
$S_T(z,z^*)=\arg\sigma(z)+\text{Im}(\alpha{z}^2/2)$, where the Weierstrass
$\sigma$-function obeys $\partial_z\sigma(z)=\zeta(z)\sigma(z)$.

In the following, we will determine the numerical solution of the GP equation
in the periodic Tkachenko gauge, Eq.~(\ref{eq:periodicGP}), for arbitrary
values of $g$.
Since the ground state solution of Eq.~(\ref{eq:periodicGP}) can acquire a
non-vanishing phase, the velocity field associated with $\psi$ does not
coincide in general with $\bm{v}_T$, although the deviations are always found
to be very small.
Actually, the velocity field coincides with $\bm{v}_T$ not only in the
incompressible regime ($g\to\infty$) but also in the LLL limit ($g\to0$).
Indeed the LLL wave function can be written in the form \cite{M-H spin}
\be
\psi_{\text{LLL}} = \prod_{jk}(z-z_{jk})e^{-|z|^2/2} =
\sigma(z)e^{\alpha{z}^2/2}e^{-|z|^2/2}
\ee
and hence, by using the explicit expression for $S_T$ reported above, one
remarkably gets $\psi_{\text{LLL}}=|\psi_{\text{LLL}}|e^{iS_T}$.

We have solved Eq.~(\ref{eq:periodicGP}) by restricting the calculation to the
parallelogram given by the lattice vectors (see Fig.~\ref{fig:1}(b)) and
imposing periodic boundary conditions.
After discretizing the continuous problem on a triangular mesh, we used the
finite element method \cite{ClaesJohnson} to find solutions by means of
imaginary time propagation.
The energies of different lattices are calculated by
varying the lattice vectors at constant cell area.
We verified that the triangular geometry (where $\alpha=0$) is the
favourite one for any value of $g>0$, while for $g=0$ any lattice with
$A_{\text{cell}}=\pi$ has the same energy, reflecting the infinite LLL
degeneracy.
For $g<0$ we find that the system is unstable, as evident from its negative
compressibility.
In Fig.~\ref{fig:1} the equilibrium density profile extracted from the
numerical solution of Eq.~(\ref{eq:periodicGP}) for $g\to0^+$ and $g=300$ is
presented. Vortices are characterized by holes in the density, whose size
increases by decreasing the interaction.
When $g\gg1$, the density is uniform everywhere except in a very narrow region
of size much smaller than the inter-vortex distance and the TF approximation
applies.
On the contrary, if $g\ll1$ the vortex core radius becomes comparable to
$l_\Omega$ and the density profile is highly structured
(density maxima and saddle points alternate on the boundary of the hexagonal
Wigner-Seitz cell).

\begin{figure}
\includegraphics[width=8.1cm]{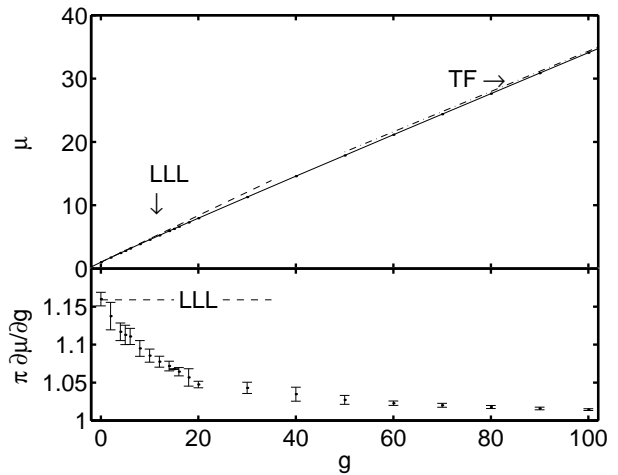}
\caption{\label{fig:2}%
Upper panel: numerically computed chemical potential $\mu$ (solid line)
as a function of $g$. The dashed and dash-dotted lines are the LLL and TF
predictions respectively.
Lower panel: derivative $\pi\partial\mu/\partial{g}$.
Numerical results (points with error bars) reproduce the correct
behaviour in both the LLL and TF regimes.}
\end{figure}

It is also interesting to study the behaviour of the chemical potential $\mu$
as a function of $g$. Making a perturbative expansion around the LLL triangular
lattice, one can prove that $\mu$ is linear in $g$ for $g\to0$, i.e., $\mu\sim
1+\beta{g}/\pi$ \cite{zero energy} with
$\beta=\int|\psi_{\text{LLL}}|^4/(\int|\psi_{\text{LLL}}|^2)^2=1.1596$
\cite{aftalion}.
By using our numerical solution, the equation of state $\mu=\mu(g)$ can be
calculated for any value of $g$, providing the bridge between the $g=0$ result
and the TF relation $\mu=g/\pi$, valid for $g\gg 1$. Since
these two opposite limits both exhibit a linear dependence with a
different slope, the transition region shows a non linear
behaviour. Deviations from linearity are however very small. The results are
reported in Fig.~\ref{fig:2}, where we also plotted the derivative
$\partial\mu/\partial{g}$ to highlight the transition.

The behaviour of $\mu(g)$ ($g\propto\langle n\rangle$) can be
used to evaluate the coarse
grained density profile $\langle n\rangle$ of the trapped
rotating gas in terms of the residual effective potential
$V_\text{eff}$, through the local density relationship
$\mu(\langle n\rangle)+V_\text{eff}=\text{const}$. In
particular, for the experimentally relevant case of harmonic
trapping, one notices that the quasi-linear behaviour of
$\mu(g)$ implies the density profile to be an inverted parabola \cite{parabola}.

\begin{figure}
\includegraphics[width=8.1cm]{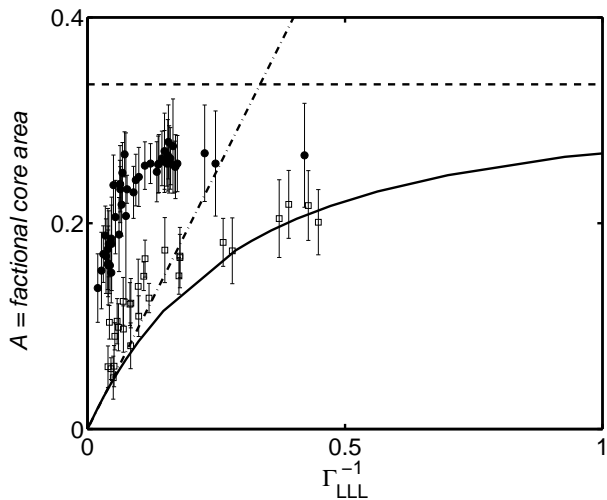}
\caption{\label{fig:3}%
Fractional vortex core area $\mathcal{A}$ as function of $\Gamma_{LLL}^{-1}$.
Solid line: numerical solution of Eq.~(\ref{eq:periodicGP}).
Dashed line: LLL ($\Gamma_{LLL}^{-1}\to\infty$) limit from
Eq.~(\ref{eq:periodicGP}).
Dash-dotted line: TF ($\Gamma_ {LLL}^{-1}\to0$) limit
$\mathcal{A}\simeq4\xi^2$.
Points with error bars: experimental results from Ref.~\protect\cite{JILA} (see
text).}
\end{figure}

Another important quantity to study is the vortex core size. In
the TF limit this length scale is fixed by the healing length
$\xi=1/\sqrt{2\mu}$, while in the LLL limit, where $\xi$ diverges,
the core size saturates to a fraction of the inter-vortex distance
\cite{F-B,B-P}. Several definitions of the vortex core radius
$r_v$ are possible. For a detailed comparison with experiments, we
follow the procedure of Ref.~\cite{JILA}, where $r_v$ is
defined as the mean square root radius of the Gaussian which
better fits the function $n_{\text{max}}-n(\bm{r})$,
evaluated on a single lattice cell \cite{r_v def}.
Our numerical results are reported in Fig.~\ref{fig:3}, where we
plot the fractional core area
$\mathcal{A}=\pi{r}_v^2/A_{\text{cell}}={r}_v^2$ as a function of the
inverse of the LLL parameter $\Gamma_{\text{LLL}}=(\mu-1)/2$ \cite{JILA}.
For $\Gamma_{\text{LLL}}\gg1$ the vortex core radius follows the TF
behaviour ($r_v\simeq1.98\xi$), while for $\Gamma_{\text{LLL}}\to0$ it
reaches the limiting value $\mathcal{A}=0.34$. The latter result is not far
from the value $\mathcal{A}=0.30$ derivable using the Gaussian fit definition
with the circular cell approximation of Ref.~\cite{B-P}. 

Concerning the comparison with experimental results \cite{JILA},
two data sets are presently available. They differ in the
procedure employed for the expansion, which is required to reach
a sufficient resolution in the imaging process. The first set
(open squares in Fig.~\ref{fig:3}) gives the fractional core area as
measured during an expansion which preserves the vertical size, while the
second one (filled circles) is obtained allowing also for some
axial expansion. Being similar to a 2D scaling,
the first procedure is expected to preserve the initial value of the
fractional area $\mathcal{A}$ \cite{2D scaling} and indeed the corresponding
experimental data are in good agreement with theory. The second technique,
involving an additional axial expansion, brings instead $\mathcal{A}$ closer to
the LLL limiting value, since it lowers the value of the effective 2D
coupling constant.

\begin{figure}
\includegraphics[width=8.5cm]{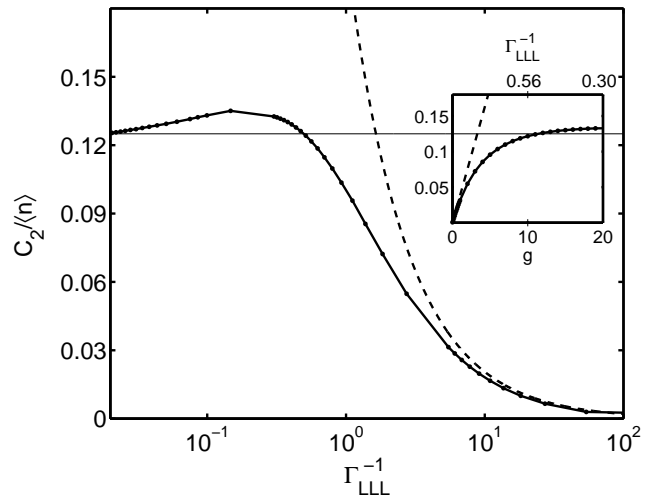}
\caption{\label{fig:4}%
Shear modulus $C_2/\langle{n}\rangle$ as a function of $\Gamma_ {LLL}^{-1}$.
Points connected by the solid line: numerical solution of
Eq.~(\ref{eq:periodicGP}).
Dashed line: LLL prediction of Ref.~\protect\cite{C2Sonin}.
Thin solid line: TF prediction of Ref.~\protect\cite{Baym 1983}.
In the inset the same data are plotted as a function of $g$ (bottom label)
and $\Gamma_ {LLL}^{-1}$ (upper label).
}
\end{figure}

We finally studied the elastic shear modulus $C_2$. This quantity, related to
the energy variation due to a shear distortion of the lattice \cite{Baym 1983},
is a crucial ingredient for the determination of the frequencies of Tkachenko
modes \cite{tka modes,C2Baym,C-P-S,C2Sonin,Sinova}. In order to extract this
coefficient, we calculate the energy for different lattice shapes, starting
from the triangular lattice and then varying one of the lattice vectors at
fixed cell area: $\omega_2'\to\omega_2+\epsilon\omega_1$. The coefficient $C_2$
is then given by $C_2=3\delta\mathcal{E}/4\epsilon^2$, where
$\delta\mathcal{E}$ is the (second order) variation of the energy per lattice
cell.

Our numerical results are reported in Fig.~\ref{fig:4} as a function of
$\Gamma_{\text{LLL}}$. In the TF limit $\Gamma_{\text{LLL}},g\gg1$ we recover
the result $C_2/\langle{n}\rangle=1/8$ \cite{Baym 1983}, while in the LLL limit
we find perfect agreement with the prediction of Refs.~\cite{Sinova,C2Sonin},
$C_2/\langle{n}\rangle\sim0.1191g$, which is a factor 10 larger than the value
previously estimated in Ref.~\cite{C2Baym}.
The figure shows that the transition between
the two regimes is characterized by a maximum at $\Gamma_{\text{LLL}} \simeq 5$. A
consequence of this non monotonic behaviour is that the LLL regime of $C_2$ is
reached at smaller values of $\Gamma_{\text{LLL}}$ with respect to the case of the vortex
core (see Fig.~\ref{fig:3}). The results for the elastic shear modulus are
relevant for the interpretation of the experiments of Ref.~\cite{JILA LLL} on the
Tkachenko oscillations in harmonically trapped condensates. Indeed, one can
average the predicted bulk values for $C_2$ over the TF profile of the trapped
condensate, using a local density approximation (see for example Ref. \cite{C-P-S}).
This procedure allows to estimate the Tkachenko frequencies also in the
intermediate regime between the TF and the LLL limit.
In practice, for the highest rotation rate realized in the experiments,
$\Omega/\omega_\perp=0.99$, where the gas is in a 2D regime and $\omega_\perp$
is the radial trapping frequency, we find that $C_2$ is lowered by only about
30\% with respect to the TF value $\langle{n}\rangle/8$. This means that the
experimental data are still far from the LLL limit of the Tkachenko frequency,
where the shear modulus would depend linearly on $g$ \cite{C2Sonin}.
Furthermore, the predicted reduction of $C_2$ is
not sufficient to explain the sizable discrepancy between the experimental
measurements and the TF prediction \cite{C-P-S}.
Possible explanations of the remaining discrepancies are:
(i) inadequacy of the local density approximation in the calculation of the
Tkachenko frequencies,
(ii) effects associated with the strong observed damping,
(iii) deviations from linearity in the experimental excitation of vortex modes,
and (iv) occurrence of anharmonic effects in the trapping potential.

In conclusion, we have investigated the bulk properties of a
rotating condensate in the presence of exact compensation between
the centrifugal potential and the harmonic trapping.
To this purpose, the mean field GP equation has been solved
by introducing a gauge transformation which explicitly exploits
the periodic structure of the density distribution and of the velocity
field in the rotating frame.
We have systematically investigated the role of interactions exploring the
transition between the Thomas-Fermi and the LLL regime.
We have found that, while the size of the vortex core exhibits a smooth
monotonic behaviour as a function of the interaction parameter, the elastic
shear modulus is characterized by an intermediate maximum.
The comparison with
experiments reveals a good agreement as concerns the size of the vortex
core and the static properties of the lattice, whereas we proved that
the remaining discrepancies regarding the Tkachenko oscillations
measured at the highest angular velocities cannot be attributed to
LLL effects alone.

\begin{acknowledgments}

We are indebted with L.P.~Pitaevskii, A.L.~Fetter, J.~Dalibard, and
N.H.~Lindner for fruitful discussions. We would
like to thank E.~Cornell and V.~Schweikhard for valuable comments and
sharing their data. C.T. would like to warmly
thank M.~Paolini for introducing him to the finite element method.

\end{acknowledgments}


\begin{thebibliography}{99}
\bibitem{2D}
By considering a system uniform along the $z$-axis with a vertical
length $Z$, the effective 2D coupling constant is given by $g_{2D}=g_{3D}/Z$,
where $g_{3D}$ is the 3D coupling constant.
If instead the motion along $z$ is squeezed into the lowest oscillator state,
one has $g_{2D}=g_{3D}/a_z\sqrt{2\pi}$, where $a_z$ is the axial oscillator
length.
\bibitem{complex}
Such description proves to be extremely useful whenever
$\divr\bm{v}=\curl\bm{v}=0$. Indeed, this implies the condition
$\partial_z{v}=0$, i.e., $v$ turns out to be an analytic function of $z^*$. For
example, for the particularly simple case of a single vortex, one has
$v=i/z^*$.
\bibitem{zero energy}
For $g=0$ one has $\mu=1$, corresponding to the zero point energy of the
2-dimensional harmonic trapping.
\bibitem{parabola}
The parabolic shape of the coarse grained density profile has been recently
proven in the LLL regime starting from the exact solution of the GP equation
\cite{watanabe,aftalion}.
\bibitem{r_v def}
In Ref.~\cite{B-P} $r_v$ is instead defined as the mean square root radius of
$n_{\text{max}}-n$ on a single cell. However, such
definition is not suitable for the TF regime, where it fails in reproducing
the proportionality between $r_v$ and $\xi$.
\end{thebibliography}
\end{document}